\documentclass[aps,prl,twocolumn,showpacs,preprintnumbers]{revtex4}
\usepackage{eurosym}
\usepackage{amsmath}
\usepackage{dcolumn}
\usepackage{bm}
\usepackage{subfigure}
\usepackage{amsfonts}
\usepackage{amssymb}
\usepackage{makeidx}
\usepackage{epsfig}
\usepackage{graphicx}

\begin{document}

\title{Microscopic statistical description of incompressible Navier-Stokes
granular fluids}
\author{Massimo Tessarotto}
\affiliation{Department of Mathematics and Geosciences, University of Trieste, Via
Valerio 12/1, 34127 Trieste, Italy}
\affiliation{Institute of Physics, Faculty of Philosophy and Science, Silesian University
in Opava, Bezru\v{c}ovo n\'{a}m.13, CZ-74601 Opava, Czech Republic}
\author{Michael Mond}
\affiliation{Department of Mechanical Engineering, Ben Gurion University of the Negev,
Be'er Sheva, Israel}
\author{Claudio Asci}
\affiliation{Department of Mathematics and Geosciences, University of Trieste, Via
Valerio 12/1, 34127 Trieste, Italy}
\date{\today }

\begin{abstract}
Based on the recently-established Master kinetic equation and related Master
constant H-theorem which describe the statistical behavior of the
Boltzmann-Sinai classical dynamical system for smooth and hard spherical
particles, the problem is posed of determining a microscopic statistical
description holding for an incompressible Navier-Stokes fluid. The goal is
reached by introducing a suitable mean-field interaction in the Master
kinetic equation. The resulting Modified Master Kinetic Equation (MMKE) is
proved to warrant at the same time the condition of mass-density
incompressibility and the validity of the Navier-Stokes fluid equation. In
addition, it is shown that the conservation of the Boltzmann-Shannon entropy
can similarly be warranted. Applications to the plane Couette and Poiseuille
flows are considered showing that they can be regarded as final decaying
states for suitable non-stationary flows. As a result, it is shown that an
arbitrary initial stochastic $1-$body PDF evolving in time by means of MMKE
necessarily exhibits the phenomenon of Decay to Kinetic Equilibrium (DKE),
whereby the $1-$body PDF asymptotically relaxes to a stationary and
spatially-uniform Maxwellian PDF.
\end{abstract}

\maketitle




\section{1 - Introduction}

Statistical approaches to the Incompressible Navier-Stokes Equations (INSE)
usually adopt one of the following routes:

1)\emph{\ Asymptotic approach} \emph{adopting high-Knudsen number and
suitable slow-velocity asymptotic approximations of the Boltzmann kinetic
equation} (see for example Refs. \cite%
{CHAPMA-COWLING,CERCIGNANI,BARDOS1991,Belardinelli2014});

2) \emph{Asymptotic approach based on so-called Lattice-Boltzmann Methods},
\textit{i.e.}, adopting suitable discrete-velocities approximation schemes
for the Boltzmann kinetic equation (see for instance Refs. \cite%
{MCNAMARA,HIGUERA1988,succi2002,CHEN1992});

3) \emph{Non-asymptotic approach} \emph{based on the so-called Inverse
Kinetic Theory }(IKT, see Refs. \cite%
{ikt1,ikt2,ikt3,ikt4,ikt5,ikt6,ikt7,ikt8}). This is based on the adoption of
a mean-field interaction acting on a set of collisionless point particles
described by means of the Vlasov kinetic equation.

In all cases indicated above the underlying dynamical system is usually
considered deterministic and time-reversible, despite the fact that the
theory may exhibit in some sense a property of macroscopic irreversibility.
In the case of the Boltzmann equation this is due to the celebrated namesake
H-theorem, while for INSE this arises because of the irreversible behavior
produced by a finite viscosity acting on the Navier-Stokes equation.
Nevertheless, the difference between these two approaches is notable. In
fact, in contrast to the IKT-approach, it is well known that Boltzmann
H-theorem is usually interpreted as being due to the phenomenon of decay to
kinetic equilibrium, namely to the occurrence of an irreversible behavior
for the kinetic probability density function (PDF) itself. \ Departing from
these views, in accordance to the GENERIC statistical model proposed by
Grmela and Oettinger (see Refs. \cite{Grmela1997,Grmela1997-2}), the
underlying dynamical system should be, instead, time-irreversible too (\emph{%
microscopic irreversibility}).

The issue arises, however, whether it is possible to reconcile the two
approaches 1) and 3) indicated above, namely to formulate a statistical
description for INSE which has the following features:

\begin{itemize}
\item first, it is non-asymptotic;

\item second, the underlying classical dynamical system remains reversible
across arbitrary instantaneous collision events while exhibiting at the same
time a possible irreversible behavior due to the mean-field force acting on
the individual particles;

\item third, it may exhibit the phenomenon of decay to kinetic equilibrium,
\textit{i.e.}, the irreversible time-evolution of the related kinetic PDF.
\end{itemize}

As a part of a systematic investigation on the statistical description of
granular fluids \cite{noi1,noi2,noi3,noi4,noi5,noi6,noi7}, identified here
either with finite or large ensembles of finite-size extended particles
(namely in which the number of particles $N$ is respectively considered of
order $O(1)$ or $\gg 1$) , the aim of the paper is to look for a possible
realization of a microscopic statistical description for INSE. In
particular, the problem is formulated in the case of the so-called
Boltzmann-Sinai classical dynamical system (CDS), or $S_{N}-$CDS \cite{noi1}%
), which advances in time an ensemble of $N$ finite-size hard-spheres
undergoing instantaneous unary, binary or multiple elastic collisions and is
also subject to the action of a suitably-prescribed external mean-field
interaction.

\subsection{Motivations and open problems}

A key issue in fluid dynamics is the identification of the appropriate
continuous fluid equations, if they exist at all, for granular systems,
namely discrete ensembles of finite-size particles which are subject to
mutual binary and/or multiple collisions as well as external interactions.
Examples are ubiquitous including: \emph{Example \#1:} \emph{Environmental}
and \emph{material-science granular fluids} (ambient atmosphere, sea-water
and ocean dynamics, etc.); \emph{Example \#2:} \emph{Biological granular
fluids} (bacterial motion in fluids, cell-blood dynamics in the human body,
blood-vessels, capillaries, etc.); \emph{Example \#3:} \emph{Industrial
granular fluids} (grain or pellet dynamics in metallurgical and chemical
processes, air and water pollution dynamics, etc.); \emph{Example \#4:}
\emph{Geological fluids} (slow dynamics of highly viscous granular fluids,
inner Earth-core dynamics, etc.). In most of the cases indicated above it is
well known that a consistent statistical description, and in particular a
fluid one, is still missing or remains largely unsatisfactory to date.

Crucial aspects involve actually the following key aspects: \emph{A) first
requirement:} the proper prescription of the dynamics of granular particles,
to be intended as classical particles subject to suitable unary, binary and
multiple interactions; \emph{B) the second requirement:} their so-called
microscopic statistical description based on classical statistical
mechanics, which involves the prescription of a suitable Liouville equation
which determines the time evolution of the corresponding phase-space PDF;
\emph{C) the third requirement} is to seek a possible finite set of
continuous fluid fields \ $\{Z(\mathbf{r},t)\}\equiv \left\{ Z_{i}(\mathbf{r}%
,t),\text{ }i=1,..,k\right\} $ which uniquely identify the "macroscopic"
state to be associated with the same granular system and satisfy identically
a closed set of PDE's, \textit{i.e.}, a finite system of differential
equations which are referred to as \emph{fluid equations,}%
\begin{equation}
F_{j}(Z(\mathbf{r},t),\mathbf{r},t)=0,  \label{fluid equatiions}
\end{equation}%
with $F_{j}$ for $j=1,..n,$ suitably-smooth real functions. In particular, $%
Z_{i}(\mathbf{r},t),$ for $i=1,..,k$ represent continuous real tensor fields
defined on the set $\Omega \times I$, with $\Omega \subset
\mathbb{R}
^{3}$ the configuration space (\emph{fluid domain}), \textit{i.e.}, either a
bounded or unbounded open and connected subset of the real Euclidean space
and $I\equiv
\mathbb{R}
$ the real time axis. \

The possibility of actually achieving a closed set of PDE's of this type
depends, however, critically on the realization of the requirements
indicated above. In this paper we intend to show that such a goal can
reached by suitably prescribing a mean-field interaction $\mathbf{F}$ acting
on the individual granular particles while still taking into account the
mutual interactions occurring among different particles. In particular, we
intend to show that in this way a fluid description of granular fluids can
be achieved in terms of the so-called incompressible Navier-Stokes equations
(INSE). More, precisely this requires, first, identifying $\{Z(\mathbf{r}%
_{1},t)\}$ with the so-called Navier-Stokes fluid fields $\{Z(\mathbf{r}%
_{1},t)\}\equiv \{\rho ,\mathbf{V},p\},$ with $\rho (\mathbf{r}_{1},t)\geq 0,%
\mathbf{V}(\mathbf{r}_{1},t)$ and $p(\mathbf{r}_{1},t)$ representing
respectively the mass density, the fluid velocity and the fluid pressure.
Notice, in particular, that here the fluid pressure is allowed for greater
generality to take also negative values. These fluid fields are assumed to
be defined and of class $C^{2}$ in the open set $\Omega $ $\times I,$ with $%
\mathbf{r}_{1}$ spanning the configuration space $\Omega ,$ to be identified
with a connected open subset of the Euclidean space $%
\mathbb{R}
^{3},$ and $t$ belonging to the oriented real time axis $I\equiv
\mathbb{R}
^{+}.$ In particular in the same open set $\Omega $ $\times I$ both $\rho (%
\mathbf{r}_{1},t)$ and $p(\mathbf{r}_{1},t)$ are assumed strictly positive.
Second, the same fluid fields are required to satisfy in same domain a
suitable initial and boundary-value problem associated with the set of fluid
equations denoted as \emph{incompressible Navier-Stokes equations }(INSE),
namely respectively%
\begin{gather}
\nabla _{1}\cdot \mathbf{V}=0,  \label{INSE-1} \\
\rho _{o}\frac{\partial }{\partial t}\mathbf{V}+\rho _{o}\mathbf{V}\cdot
\nabla _{1}\mathbf{V}+\nabla _{1}p-\mathbf{f}-\mu \nabla _{1}^{2}\mathbf{V}%
=0,  \label{INSE-2}
\end{gather}%
where in particular in the open set $\Omega $ the initial conditions are of
the form
\begin{equation}
\left\{
\begin{array}{c}
\rho (\mathbf{r}_{1},t_{o})=\rho _{o}, \\
\mathbf{V}(\mathbf{r}_{1},t_{o})=\mathbf{V}_{o}(\mathbf{r}_{1}), \\
p(\mathbf{r}_{1},t_{o})=p_{o}(\mathbf{r}_{1}),%
\end{array}%
\right.  \label{INITIAL COND}
\end{equation}%
with $\rho _{o}>0$ a constant mass density, $p_{o}(\mathbf{r}_{1})$ the
initial fluid pressure and $\mathbf{V}_{o}(\mathbf{r}_{1})$ the
corresponding initial fluid velocity which by assumption must fulfill the
isochoricity equation (\ref{INSE-1}). Then by construction it follows that
in the domain $\Omega \times I$ the fluid pressure takes the form%
\begin{equation}
p(\mathbf{r}_{1},t)=\int\limits_{\Omega }d^{3}\mathbf{r}_{1}^{\prime }\frac{%
S(\mathbf{r}_{1}^{\prime },t)}{\left\vert \mathbf{r}_{1}\mathbf{-r}%
_{1}^{\prime }\right\vert }+p_{f},  \label{POISSON}
\end{equation}%
with $p_{f}$ here assumed to be an in principle arbitrary real constant$.$
Moreover, requiring that the force density field $\mathbf{f}$ to be
divergence-free and such that $\mathbf{f}$, $S(\mathbf{r}_{1}^{\prime },t)$
is the source term%
\begin{equation}
S(\mathbf{r}_{1},t)=-\rho _{o}\nabla _{1}\mathbf{V:}\nabla _{1}\mathbf{V.}
\end{equation}%
In addition, Dirichlet boundary conditions of the form%
\begin{equation}
\left\{
\begin{array}{c}
\left. \rho (\mathbf{r}_{1},t)\right\vert _{\mathbf{r}_{1}\in \partial
\Omega }=\rho _{o}, \\
\left. \mathbf{V}(\mathbf{r}_{1},t)\right\vert _{\mathbf{r}_{1}\equiv
\mathbf{r}_{w}\in \partial \Omega }=\mathbf{V}_{w}(\mathbf{r}_{w}),%
\end{array}%
\right.  \label{BOUNDARY COND}
\end{equation}%
are considered with $\mathbf{V}_{w}(\mathbf{r}_{w})$ a suitably smooth
vector function, while $\left. p_{1}(\mathbf{r}_{1},t)\right\vert _{\mathbf{r%
}_{1}\equiv \mathbf{r}_{w}\in \partial \Omega }$ is considered uniquely
prescribed by Eq. (\ref{POISSON}). Here the notation is standard. Thus, $%
\nabla _{1}$ is the gradient operator $\nabla _{1}\equiv \frac{\partial }{%
\partial \mathbf{r}}.$ The first and second equations are known respectively
as the so-called \emph{isochoricity }and \emph{Navier-Stokes} equations,
while $\mathbf{f},\mu >0$ and $\rho (\mathbf{r}_{1},t)$ identify
respectively a suitable volume force, the constant fluid viscosity and the
mass density. The latter is assumed constant in the set $\overline{\Omega }%
\times I,$ with $\overline{\Omega }$ being the closure of $\Omega ,$\textit{%
i.e.},%
\begin{equation}
\rho (\mathbf{r}_{1},t)=\rho _{o}>0,  \label{incomp0ressibility cond}
\end{equation}%
the same equation being referred to as \emph{incompressibility condition}.
In addition we shall introduce for $\mathbf{f}$ the decomposition%
\begin{equation}
\mathbf{f=f}_{1}+\mathbf{f}_{2},  \label{DECOMPOSITION}
\end{equation}%
where respectively $\mathbf{f}_{1}$ and $\mathbf{f}_{2}$ are assumed of the
form $\mathbf{f}_{1}=\mathbf{f}_{1}(\mathbf{r}_{1},t)\equiv \mathbf{-\nabla }%
_{1}\phi _{1}$ and $\mathbf{f}_{2}=\mathbf{f}_{2}(\mathbf{r}_{1})\equiv
\mathbf{-\nabla }_{1}\phi _{2},$ namely are respectively assumed
non-stationary and stationary$.$ The task posed in this paper involves
representing the same fluid fields in terms of suitable statistical averages
- \textit{i.e.}, velocity moments - to be evaluated in terms of an
appropriate phase-space PDF. More precisely, as shown below, the goal can be
realized by means of a microscopic statistical description based on the
Master kinetic equation recently developed (see Refs.\cite%
{noi1,noi2,noi3,noi4,noi5,noi6,noi7}). This refers, in particular, to the
phase-space dynamics of smooth hard-spheres whose time-evolution is
generated by the so-called Boltzmann-Sinai classical dynamical systems
(briefly denoted as $S_{N}-$CDS). In addition, as indicated above, the same
particles are assumed to be subject to the action of suitable mean-field
interactions acting on the center of mass of each particle.

In this paper the problem is posed of formulating a novel statistical
description for INSE for granular fluids\emph{\ in conditions of arbitrary
diluteness}. Although such an approach can in principle apply to a variety
flows with different kinds of grains which are in turn also possibly
embedded in another fluid (see \emph{Examples \#1-\#4} above), in the
following the attention will be focused on the kinetic treatment of a
granular system formed by a single species of smooth hard spheres undergoing
instantaneous and purely elastic collisions. Starting point is the axiomatic
treatment recently developed for the statistical description of the
Boltzmann-Sinai CDS \cite{noi1,noi2,noi3,noi4,noi5,noi6,noi7}. The present
study follows from two previous theoretical developments. The first one is
related to the establishment of statistical descriptions based on so-called
inverse kinetic theory (IKT). It relies on the axiomatic introduction of a
Liouville equation describing the collisionless dynamics of particles
subject to the action of suitable external mean-field forces (see Refs. \cite%
{ikt1,ikt2,ikt3,ikt4,ikt6}).The second development is even more significant.
It lays in the discovery of a new kinetic equation realized by Master
equation \cite{noi3}, \textit{i.e.}, an exact, namely non-asymptotic,
kinetic equation which advances in time an arbitrary ensemble of smooth
hard-sphere particles which are advanced in time by means of the $S_{N}-$CDS
indicated above. In particular, the Master equation is peculiar since it
holds for a finite system $S_{N}-$CDS, \textit{i.e.}, a finite number of
smooth hard spheres which are in turn considered as finite-size, namely
having a finite mass $m$ and a finite diameter $\sigma >0$. As discussed in
Refs.\cite{noi1,noi2,noi3,noi4,noi5,noi6} such features imply a radical
change of viewpoint in kinetic theory. However, the introduction of
mean-field (\textit{i.e.}, external) forces into the Master kinetic
equation, a task which is carried out in this paper, leads in principle to a
novel form of the same equation which is referred to here as \emph{modified
Master kinetic equation }(MMKE)\emph{. }\ It is obvious that for such an
equation the very validity of previous conclusions \cite{noi7} is called
into question. Therefore a number of basic related issues arise which
concern in particular:

\emph{Problem \#1:} The first one is whether the incompressibility condition
can achieved or not. This is realized by the requirement
\begin{equation}
n(\mathbf{r}_{1},t)=n_{o},
\end{equation}%
with $n(\mathbf{r}_{1},t)$ denoting the Navier-Stokes configuration-space
probability density and $n_{o}>0$ a constant value to hold in the interior
fluid domain $\Omega $.

\emph{Problem \#2:} The second issue concerns the validity itself of the
Navier-Stokes equation, \textit{i.e.}, whether it holds identically in $%
\Omega \times I$. This means, in particular, that in the absence of external
volume forces, except for gravity, its intrinsic time-irreversible character
should be warranted (\emph{macroscopic irreversibility}).

\emph{Problem \#3:} The third one concerns the possible validity of a
constant H-theorem holding also for the modified Master kinetic equation, in
terms of the Boltzmann-Shannon statistical entropy associated with the
kinetic PDF.

\emph{Problem \#4: }The final issue is whether an intrinsic macroscopic
irreversibility phenomenon can occur for the modified Master kinetic
equation. More precisely, this is related to the possible occurrence of the
decay to kinetic equilibrium (DKE) for the $1-$body PDF in the limit $%
t\rightarrow +\infty $, \ i.e., in which the same PDF coincides with a
stationary solution of the modified Master kinetic equation.

\subsection{Goals of the investigation}

Given the premises indicated above, the goals to be pursued in the paper are
as follows:

\emph{GOAL \#1 (Section 2): Review of the Master kinetic equation and its
basic properties.}

\emph{GOAL \#2 (Section 3): The introduction of the modified Master kinetic
equation (MMKE). }This concerns also the establishment of the related moment
equations and the related evolution equation for the Boltzmann-Shannon
entropy.

\emph{GOAL \#4 (Section 3, Subsection 3.1): The problem of incompressibility
for MMKE. }The goal here is to prescribe the a suitable form of the
mean-field interaction in such a way to fulfill identically in $\Omega
\times I$ the incompressibility condition.

\emph{GOAL \#5 (Section 3, Subsection 3.2): The problem of Navier-Stokes
equation for MMKE. }We intend to investigate the possible realization of the
Navier-Stokes equation based on \emph{\ MMKE}.

\emph{GOAL \#6 (Section 3, Subsection 3.3): The problem of the entropy
conservation for MMKE. } In this subsection the constant H-theorem is
established.

\emph{GOAL \#7 (Section 4): Example applications of MMKE: the plane Couette
and the Poiseuille flows}. The\ goal here is to test the validity of the
modified Master kinetic equation to describe particular stationary solutions
of INSE.

\emph{GOAL \#8 (Section 5): Physical implications of the theory. }These
concern the possible occurrence of the phenomenon of DKE for MMKE.

\section{2 - The Master kinetic equation and the Master H-theorem}

Although Boltzmann's namesake equation and Boltzmann H-Theorem \cite%
{Boltzmann1972,Boltzmann1896,Boltzmann1896c,Boltzmann1897} still represent
nowadays a historical breakthrough, certain aspects of their derivation and
in particular their extension to the treatment of granular systems,\textit{\
i.e.}, formed by finite-size particles, have remained for a long time
unsatisfactory \cite{Enskog,noi3}. A critical issue in this connection is
the physical basis of the involved microscopic statistical description \cite%
{Grad,noi1}. This refers in particular to the unique identification of the
correct physical prescriptions - at arbitrary collision events - for the
time-evolution laws of the $s-$body (for $s=2,..,N$) probability density
functions (PDF) associated with the Boltzmann-Sinai S$_{N}-$CDS namely a
system formed by a finite number $N$ of finite-size smooth hard spheres of
diameter $\sigma $ and mass $m$ undergoing unary, binary as well in
principle arbitrary multiple instantaneous elastic collisions \cite{noi7}.
More precisely, the issue is to determine the relationship between the $s-$%
body PDF's occurring immediately before and after an instantaneous collision
event, \textit{i.e.},, the so-called incoming and outgoing PDF's $\rho
_{s}^{(N)(-)}$ and $\rho _{s}^{(N)(+)}$ (with $s=1,..,N$)$,$ referred to as
\emph{collision boundary condition}$.$(CBC) \cite{noi1} for the $s-$body
reduced PDF $\rho _{s}^{(N)}$ which is assumed function of the corresponding
$s-$body state $\mathbf{x}^{(s)}\equiv \left\{ \mathbf{x}_{1},...,\mathbf{x}%
_{s}\right\} $ and time $t.$ Here for definiteness, $\mathbf{x}_{i}\equiv
\left\{ \mathbf{r}_{i},\mathbf{v}_{i}\right\} $ for $i=1,N$ denotes the $i-$%
the particle state of the S$_{N}-$CDS belonging to the corresponding $1-$%
body phase space $\Gamma _{1}\equiv \Omega \times U_{1}.$ In addition, $%
\mathbf{r}_{i}$ and $\mathbf{v}_{i}$ label its center of mass position and
velocity. The latter are assumed to span the configuration and velocity
spaces, here identified respectively with the fluid domain $\Omega $ and $%
U_{1}\equiv \mathbb{R} ^{3}.$ In this connection the following physical
prescriptions are mandatory for the determination of the appropriate CBC:

\emph{Physical prescription \#1:} \emph{causality principle} requires
representing $\rho _{s}^{(N)(+)}$ as a function of $\rho _{s}^{(N)(-)},$
thus predicting the future from the past (rather than the opposite). Such a
causal choice according to Cercignani \cite{Cercignani1975} determines the
arrow of time, \textit{i.e.}, the orientation of the time axis;

\emph{Physical prescription \#2: }the \emph{axiom of probability conservation%
} at arbitrary collision events must be fulfilled for appropriate subsets of
the $N-$body phase-space (see related discussion in Refs. \cite%
{noi1,noi2,noi3});

\emph{Physical prescription \#3:} the \emph{local} \emph{prescription }of
the appropriate collision boundary conditions must be adopted. More
precisely, this means that CBC should be realized by means of a \emph{local
relationship} between $\rho _{s}^{(N)(+)}$ and $\rho _{s}^{(N)(-)}$ when
both are evaluated \emph{at the same state }belonging to the $s-$body
phase-space $\Gamma _{s}$. In other words, this requires prescribing the
functional form of the outgoing PDF in terms of the same outgoing state,%
\textit{\ i.e.}, after collision. The latter one can be equivalently
expressed in Lagrangian or Eulerian states, namely either in terms of $%
\mathbf{x}^{(s)(+)}(t_{k})$ or $\mathbf{x}^{(s)(+)}$ (with $\mathbf{x}%
^{(s)(+)}(t_{k})$ belonging to a prescribed Lagrangian trajectory $\left\{
\mathbf{x}^{(s)(+)}(t),t\in I\right\} $). Consider for definiteness the case
of a two-body collision.. Then, upon identifying for $i=1,2$, $\mathbf{x}%
_{i}^{(+)}(t_{k})\mathbf{\equiv }\left\{ \mathbf{r}_{i}(t_{k}),\mathbf{v}%
_{i}^{(+)}(t_{k})\right\} $ and $\mathbf{x}_{i}^{(-)}(t_{k})\mathbf{\equiv }%
\left\{ \mathbf{r}_{i}(t_{k}),\mathbf{v}_{i}^{(-)}(t_{k})\right\} $ with the
corresponding outgoing and incoming states and letting $\mathbf{r}%
_{2}(t_{k})\equiv \mathbf{r}_{1}(t_{k})+\sigma \mathbf{n}_{21}(t_{k}),$ this
means in particular that%
\begin{equation}
\left\{
\begin{array}{c}
\mathbf{v}_{1}^{(+)}(t_{k})=\mathbf{v}_{1}^{(-)}(t_{k})-\mathbf{n}%
_{12}(t_{k})\mathbf{n}_{12}(t_{k})\cdot \mathbf{v}_{12}^{(-)}(t_{k}), \\
\mathbf{v}_{2}^{(+)}(t_{k})=\mathbf{v}_{2}^{(-)}(t_{k})-\mathbf{n}%
_{21}(t_{k})\mathbf{n}_{21}(t_{k})\cdot \mathbf{v}_{21}^{(-)}(t_{k}).%
\end{array}%
\right.
\end{equation}%
Here the notation is standard \cite{noi1}. Thus, $\mathbf{r}%
_{12}(t_{k})\equiv \mathbf{r}_{1}(t_{k})-\mathbf{r}_{2}(t_{k}),$ $\mathbf{v}%
_{12}^{(-)}(t_{k})\equiv \mathbf{v}_{1}(t_{k})-\mathbf{v}_{2}(t_{k})$ and $%
\mathbf{n}_{12}(t_{k})=$ $\mathbf{r}_{12}(t_{k})/\left\vert \mathbf{r}%
_{12}(t_{k})\right\vert $ denote respectively the relative position and
velocity vectors and the corresponding relative position unit vector. \ As
shown in Ref. \cite{noi2} the correct causal prescription of CBC, denoted as
\emph{modified collision boundary conditions} (MCBC), is found to be
provided respectively by the Lagrangian and Eulerian equations%
\begin{equation}
\rho ^{(+)\left( N\right) }(\mathbf{x}^{\left( +\right) }(t_{k}),t_{k})=\rho
^{(-)\left( N\right) }(\mathbf{x}^{(+)}(t_{k}),t_{k}),  \label{bbb}
\end{equation}%
and%
\begin{equation}
\rho ^{(+)\left( N\right) }(\mathbf{x}^{\left( +\right) },t)=\rho
^{(-)\left( N\right) }(\mathbf{x}^{(+)},t).  \label{bbb-euler}
\end{equation}%
These prescriptions emerge clearly when the special case is considered of
the $N-$body deterministic PDF. This is realized by the $N-$body Dirac delta
(or so-called \textit{certainty function} \cite{Cercignani1969a}), namely
the distribution $\rho _{H}^{\left( N\right) }(\mathbf{x},t)=\delta \left(
\mathbf{x}-\mathbf{x}(t)\right) \equiv \prod\limits_{i=1,N}\delta (\mathbf{x}%
_{i}-\mathbf{x}_{i}(t)),$ with $\delta (\mathbf{x}_{i}-\mathbf{x}_{i}(t))$
being the $1-$body Dirac delta, $\mathbf{x}_{i}$ and $\left\{ \mathbf{x}%
_{i}(t),t\in I\right\} $ denoting respectively the $i-$th particle state
which spans the corresponding $1-$body phase space and the phase-space
trajectory of the same particle. In fact, exclusively based on physical
grounds, \textit{i.e.}, thanks to the axioms of classical statistical
mechanics \cite{noi1,noi3}, the $N-$body Dirac delta must necessarily
provide a particular possible realization for the $N-$body PDF associated
the same CDS. \ On the other hand, each single-particle Dirac delta can be
regarded as the limit function of an arbitrary, \textit{i.e.}, intrinsically
non-unique, sequence of smooth strictly positive real functions. Therefore,
it is obvious that the same collision boundary conditions should manifestly
be satisfied also by the sequence-functions themselves.

As shown in Ref. \cite{noi2} the modified collision boundary conditions
indicated above (see Eqs. (\ref{bbb}) and (\ref{bbb-euler})) differ from the
traditional prescription earlier adopted in the literature and originally
first introduced by Boltzmann (see related discussion in Refs. \cite%
{Cercignani1975}) which is realized by the so-called PDF-conserving CBC,
namely the Lagrangian requirement%
\begin{equation}
\rho ^{(+)\left( N\right) }(\mathbf{x}^{\left( +\right) }(t_{k}),t_{k})=\rho
^{(-)\left( N\right) }(\mathbf{x}^{(-)}(t_{k}),t_{k}),  \label{aaa}
\end{equation}%
of the equivalent Eulerian one%
\begin{equation}
\rho ^{(+)\left( N\right) }(\mathbf{x}^{\left( +\right) },t)=\rho
^{(-)\left( N\right) }(\mathbf{x}^{(-)},t),  \label{aaa-2}
\end{equation}%
with $\mathbf{x}^{(-)}(t_{k}),\mathbf{x}^{\left( +\right) }(t_{k})$ and $%
\mathbf{x}^{(-)},\mathbf{x}^{\left( +\right) }$ denoting again the
respectively the couples of Lagrangian and Eulerian incoming and outgoing
particle states.\ Actually Eqs. (\ref{aaa}) and (\ref{aaa-2}) are peculiar
since they assertively relate two PDF's evaluated ad different phase-space
states. Thus, manifestly violating the \emph{Physical prescription \#3}
indicated above, they should be regarded as un-physical ones. Nonetheless,
the same CBC preserve by construction the customary Boltzmann collisional
invariants (see Eq.(\ref{BOLTZMANN COILLISIONAL INVARIANTS}) below; \cite%
{noi5}). In addition, when then so-called Boltzmann-Grad limit is performed,
\textit{i.e.}, point-like particles are considered in validity of the
dilute-gas asymptotic ordering, one can show \cite{noi3,noi6,noi7} that the
two choices indicated above (represented respectively by MCBC and the
PDF-conserving CBC) actually lead to the same realization of the collision
operator in the BBGKY hierarchy \cite{noi4} and in the Boltzmann equation
\cite{noi1,noi2,noi3}.

The adoption of MCBC as well the prescription of an\emph{\ }extended
functional setting for microscopic statistical description of the
Boltzmann-Sinai CDS, lay at the basis of the new\ "\textit{ab initio}"
treatment of classical statistical mechanics (CSM), recently developed in
Refs. \cite{noi1,noi2,noi3,noi4,noi5,noi6,noi7}, enabling the treatment of
granular systems formed by a finite number of particles. As a consequence,
it is found that the $N-$body PDF now can include among its
physically-admissible realizations both stochastic (\textit{i.e.}, smoothly
differentiable ordinary functions), partially deterministic and
deterministic (\textit{i.e.}, in both cases distributions) probability
density functions \cite{noi1,noi3}. In addition, the same $\Gamma _{N}-$%
phase-space Liouville equation\emph{\ }necessarily admits among its
physically-admissible solutions also a permutation-symmetric, particular
realization of the $N-$body PDF which is factorized in terms of the
corresponding $1-$body PDF's $\rho _{1}^{(N)}(\mathbf{x}_{i}\mathbf{,}t)$
for all $i=1,N$.

Such an approach has opened up a host of exciting new developments in the
kinetic theory of granular systems.

In particular, as pointed out in Refs. \cite{noi3} and \cite{noi6}, the
really remarkable implication which follows is that the same $1-$body PDF $%
\rho _{1}^{(N)}(t)\equiv \rho _{1}^{(N)}(\mathbf{x}_{1}\mathbf{,}t)$
satisfies an equivalent exact, \textit{i.e.}, non-asymptotic kinetic
equation holding for finite values of ($N,\sigma ,m$), denoted as \emph{%
Master kinetic equation }(see Eq.(\ref{MASTER EQUATION}) below)\emph{, }%
while in addition the corresponding Boltzmann-Shannon $1-$body entropy $%
S(\rho _{1}^{(N)}(t))$ (see Eq. (\ref{B-.S ENTROPY}) below) is identically
conserved for all $t\in I\equiv
\mathbb{R}
$ (\emph{Master H-theorem equation}; see Eq.(\ref{MASTER H-THEOREM EQUATION}%
) below).

For the sake of greater clarity the two equations are briefly recalled below
together with their Boltzmann-Grad limits.

\subsection{2.1 - Master kinetic equation}

In the case of a stochastic factorized $N-$body PDF the Master kinetic
equation for the corresponding stochastic reduced $1-$body PDF can be
represented in terms of the integro-differential equation%
\begin{equation}
L_{1\left( 1\right) }\rho _{1}^{(N)}(\mathbf{x}_{1},t)=\mathcal{C}_{1}\left(
\rho _{1}^{(N)}|\rho _{1}^{(N)}\right) ,  \label{MASTER EQUATION}
\end{equation}%
where the operators $L_{1\left( 1\right) }$ and $C_{1}\left( \rho
_{1}^{(N)}|\rho _{1}^{(N)}\right) $\ identify respectively the
free-streaming operators and the Master collision operator which is
consistent both with the causality principle and MCBC. These are given
respectively by%
\begin{equation}
L_{1\left( 1\right) }=L_{1\left( 1\right) }\equiv \frac{\partial }{\partial t%
}+\mathbf{v}_{1}\cdot \frac{\partial }{\partial \mathbf{r}_{1}}.
\label{free-streaming operator}
\end{equation}%
\begin{gather}
\mathcal{C}_{1}\left( \rho _{1}^{(N)}|\rho _{1}^{(N)}\right) \equiv \left(
N-1\right) \sigma ^{2}\int\limits_{U_{1(2)}}d\mathbf{v}_{2}\int^{(-)}d\Sigma
_{21}  \notag \\
\left[ \widehat{\rho }_{2}^{(N)}(\mathbf{x}^{(2)(+)},t)-\widehat{\rho }%
_{2}^{(N)}(\mathbf{x}^{(2)},t)\right] \left\vert \mathbf{v}_{21}\cdot
\mathbf{n}_{21}\right\vert \overline{\Theta }^{\ast }.
\label{MASTER-OPERATOR}
\end{gather}%
Here the notation are standard \cite{noi3}. Thus $U_{1(k)}\equiv
\mathbb{R}
^{3}$ is the $1-$body velocity space for the $k-$th particle$,$ the symbol $%
\int^{(-)}d\Sigma _{21}$ denotes integration on the subset of the solid
angle of incoming particles namely for which $\mathbf{v}_{12}\cdot \mathbf{n}%
_{12}<0.$ Furthermore
\begin{align}
\widehat{\rho }_{2}^{(N)}(\mathbf{x}^{(2)},t)& \equiv \frac{k_{2}^{(N)}(%
\mathbf{r}_{1},\mathbf{r}_{2},t)}{k_{1}^{(N)}(\mathbf{r}_{1},t)k_{1}^{(N)}(%
\mathbf{r}_{2},t)}  \notag \\
& \rho _{1}^{(N)}(\mathbf{r}_{1},\mathbf{v}_{1},t)\rho _{1}^{(N)}(\mathbf{r}%
_{2},\mathbf{v}_{2},t)  \label{ro-2}
\end{align}%
identifies the incoming $2-$body PDF. Similarly $\widehat{\rho }_{2}^{(N)}(%
\mathbf{x}^{(2)(+)},t)$\ is the corresponding outgoing $2-$body PDF's as
determined from MCBC, namely%
\begin{align}
\widehat{\rho }_{2}^{(N)}(\mathbf{x}^{(2)(+)},t)& \equiv \frac{k_{2}^{(N)}(%
\mathbf{r}_{1},\mathbf{r}_{2},t)}{k_{1}^{(N)}(\mathbf{r}_{1},t)k_{1}^{(N)}(%
\mathbf{r}_{2},t)}  \notag \\
& \rho _{1}^{(N)}(\mathbf{r}_{1},\mathbf{v}_{1}^{(+)},t)\rho _{1}^{(N)}(%
\mathbf{r}_{2},\mathbf{v}_{2}^{(+)},t).  \label{ro-2(+)}
\end{align}%
In both equations above (\ref{ro-2}) and (\ref{ro-2(+)}) the position vector%
\emph{\ }$\mathbf{r}_{2}$ is identified with\emph{\ }$\mathbf{r}_{2}=\mathbf{%
r}_{1}+\sigma \mathbf{n}_{21},$ while $k_{1}^{(N)}(\mathbf{r}%
_{1},t),k_{1}^{(N)}(\mathbf{r}_{2},t)$ and $k_{2}^{(N)}(\mathbf{r}_{1},%
\mathbf{r}_{2},t)$ identify suitably-prescribed $1-$ and $2-$body occupation
coefficients (see the corresponding definitions reported in Ref. \cite{noi3}%
). Finally $\overline{\Theta }^{\ast }$ denotes\emph{\ }$\overline{\Theta }%
^{\ast }\equiv \overline{\Theta }\left( \left\vert \mathbf{r}_{2}-\frac{%
\sigma }{2}\mathbf{n}_{2}\right\vert -\frac{\sigma }{2}\right) ,$ with $%
\overline{\Theta }(x)$ being the strong theta function.

Then it is immediate to show that the Master collision operator admits the
standard Boltzmann collisional invariants which warrants that identically
the equations%
\begin{equation}
\int\limits_{U_{1}}d\mathbf{v}_{1}G_{i}(\mathbf{x}_{1},t)\mathcal{C}%
_{1}\left( \rho _{1}^{(N)}|\rho _{1}^{(N)}\right) =0
\label{BOLTZMANN COILLISIONAL INVARIANTS}
\end{equation}%
must hold for $G_{i}(\mathbf{x}_{1},t)=1,\mathbf{v}_{1},v_{1}^{2}$.

\subsection{2.2 - Master H-theorem equation}

Let us require that $\rho _{1}^{(N)}(t)\equiv \rho _{1}^{(N)}(\mathbf{x}%
_{1},t)$ is a stochastic $1-$body PDF solution of the Master kinetic
equation (\ref{MASTER EQUATION}) such that for the corresponding initial
condition $\rho _{1}^{(N)}(t_{o})\equiv \rho _{1}^{(N)}(\mathbf{x}%
_{1},t_{o}) $\ the Boltzmann-Shannon entropy associated with $\rho
_{1}^{(N)}(t_{o})$ is defined, namely the functional \cite%
{Boltzmann1972,Shannon}%
\begin{equation}
S(\rho _{1}^{(N)}(t_{o}))=-\int\limits_{\Gamma _{1}}d\mathbf{v}_{1}\rho
_{1}^{(N)}(\mathbf{x}_{1},t_{o})\ln \rho _{1}^{(N)}(\mathbf{x}_{1},t_{o})
\end{equation}%
exists. Then it follows that:

\begin{enumerate}
\item For all $t\in I\equiv
\mathbb{R}
,$ the Boltzmann-Shannon entropy associated with a stochastic solution of
the Master kinetic equation (\ref{MASTER EQUATION}) $\rho _{1}^{(N)}(t),$
namely%
\begin{equation}
S(\rho _{1}^{(N)}(t))\equiv -\int\limits_{\Gamma _{1}}d\mathbf{v}_{1}\rho
_{1}^{(N)}(\mathbf{x}_{1},t)\ln \rho _{1}^{(N)}(\mathbf{x}_{1},t),
\label{B-.S ENTROPY}
\end{equation}%
exists globally in time (see Ref. \cite{noi7});

\item For all $t\in I\equiv
\mathbb{R}
,$ $S(\rho _{1}^{(N)}(t))$ is such that denoting $K_{1}(\rho _{1}^{(N)}(t))$
the weighted phase-space integral of the Master collision operator
\begin{equation}
K_{1}(\rho _{1}^{(N)}(t))\equiv -\int\limits_{\Gamma _{1}}d\mathbf{v}_{1}\ln
\rho _{1}^{(N)}(\mathbf{x}_{1},t)\mathcal{C}_{1}\left( \rho _{1}^{(N)}|\rho
_{1}^{(N)}\right) ,  \label{K_1}
\end{equation}%
the entropy density $G(\mathbf{x}_{1},t)\equiv \ln \rho _{1}^{(N)}(\mathbf{x}%
_{1},t)$ is a generalized collisional invariant for the Master collision
operators (see Ref.\cite{noi5}) so that the identity%
\begin{equation}
K_{1}(\rho _{1}^{(N)}(t))\equiv 0  \label{IDENRITTY FOR K_1}
\end{equation}%
necessarily holds.

\item As a consequence, since by construction%
\begin{equation}
\frac{\partial }{\partial t}S(\rho _{1}^{(N)}(t))\equiv K_{1}(\rho
_{1}^{(N)}(t)),
\end{equation}%
i.e., $K_{1}(\rho _{1}^{(N)}(t))$ is the entropy production rate associated
with the Master collision operator, the Boltzmann-Shannon entropy is
constant in time so that necessarily the constant Master H-theorem equation%
\begin{equation}
S(\rho _{1}^{(N)}(t))=S(\rho _{1}^{(N)}(t_{o}))
\label{MASTER H-THEOREM EQUATION}
\end{equation}%
is identically fulfilled (see Ref. \cite{noi6}).
\end{enumerate}

\subsection{2.3 - Boltzmann-Grad limit}

For completeness it is convenient to recall here the relationship with
Boltzmann kinetic theory, namely the asymptotic approximation by which the
Master kinetic equation (\ref{MASTER EQUATION}) recovers the customary form
of the Boltzmann equation (see also Ref.\cite{noi4}). This involves, besides
suitable smoothness conditions on the $1-$ body PDF, invoking:

A) first, the so-called\emph{\ dilute-gas asymptotic ordering} for $N$ and $%
\sigma ,$ obtained by means of the asymptotic conditions%
\begin{equation}
\left\{
\begin{array}{c}
N\equiv \frac{1}{\varepsilon }\gg 1 \\
0<\sigma \sim O(\varepsilon ^{1/2}),%
\end{array}%
\right.  \label{Dilute gas ordering}
\end{equation}%
requiring the Knudsen number $K_{n}\equiv N\sigma ^{2}$ to become of $%
O(\varepsilon ^{0})$ (see also Ref. \cite{noi4});

B) second, \emph{the continuum limit}, or Boltzmann-Grad limit, obtained
letting%
\begin{equation}
\left\{
\begin{array}{c}
\varepsilon \equiv \frac{1}{N}\rightarrow 0^{+}, \\
K_{n}\equiv N\sigma ^{2}\rightarrow K_{n}^{(o)},%
\end{array}%
\right.  \label{continuum limit}
\end{equation}%
with $K_{n}$ denoting the Knudsen number and $K_{n}^{(o)}\sim $ $%
O(\varepsilon ^{0})$ its limit value.

As shown in Refs. \cite{noi3} and \cite{noi6} these assumptions imply that
the limit function
\begin{equation}
\rho _{1}(\mathbf{x}_{1},t)=\lim_{\varepsilon \rightarrow 0^{+}}\rho
_{1}^{(N)}(\mathbf{x}_{1},t)
\end{equation}%
obtained consistently with Eqs.(\ref{Dilute gas ordering})\ and (\ref%
{continuum limit}) by construction satisfies the Boltzmann kinetic equation
and the corresponding Boltzmann H-theorem. However, requirements A) and B)
imply treating the hard-spheres as point-like and taking the continuum limit
in such a way that the Knudsen number remains finite. Therefore this means
that, unlike the case of a granular system, \textit{i.e.}, formed by
finite-size particles, the ensemble of particles corresponds to a rarefied
gas, \textit{i.e.}, an entirely different physical system from the one
considered here.

\section{3 - Modified form of the Master kinetic equation}

The task posed in this section is to determine a suitable \emph{mean-field
interaction}, represented by a smoothly-differentiable real vector field $%
\mathbf{F,}$ occurring in the Master kinetic equation and representing a
unary (\textit{i.e.}, so-called \emph{mean-field}) interaction depending
only on the state of each individual particle. The latter is assumed to act
individually on each particle belonging to a granular system, \textit{i.e.},
an ensemble of finite-size hard spheres described by means of the
Boltzmann-Sinai CDS. The goal is to show that the vector field $\mathbf{F}$
can be prescribed in such way to achieve GOALS \#2-\#6 stated in the
introduction, \textit{i.e.}, to satisfy identically INSE and to warrant at
the same time conservation of the statistical Boltzmann-Shannon entropy
associated with the relevant $1-$body PDF. The resulting statistical
equation, obtained in terms of the Master kinetic equation recalled above (%
\ref{MASTER EQUATION}) and to be referred to as \emph{modified Master
kinetic equation} (MMKE), takes the form%
\begin{equation}
L_{1}(\mathbf{F})\rho _{1}^{(N)}(\mathbf{x}_{1},t)=\mathcal{C}_{1}\left(
\rho _{1}^{(N)}|\rho _{1}^{(N)}\right) ,  \label{MODIFIED MASTER EQ}
\end{equation}%
with $L_{1}(\mathbf{F})$ denoting the modified streaming operator%
\begin{equation}
L_{1}(\mathbf{F})=\frac{\partial }{\partial t}+\mathbf{v}_{1}\cdot \nabla
_{1}+\frac{\partial }{\partial \mathbf{v}_{1}}\cdot \left( \mathbf{F}\right)
.  \label{modified STREAMING OPERATOR}
\end{equation}%
The same vector field $\mathbf{F}$ will be assumed to be a
smoothly-differentiable real vector field of the type $\mathbf{F=F}(\mathbf{x%
}_{1},t;\rho _{1}^{(N)}),$ which depends functionally also on the $1-$body
PDF $\rho _{1}^{(N)}(t)\equiv \rho _{1}^{(N)}(\mathbf{x}_{1},t)$. Here we
intend to prescribe $\mathbf{F}$ in such a way to satisfy the following
constraint conditions: a) the first one is realized by the incompressibility
condition prescribed by Eq.(\ref{incomp0ressibility cond}); b) the second
constraint is \ provided by the validity of the Navier-Stokes equation \ref%
{INSE-2}); c) the third constraint condition is realized by the requirement
that the Boltzmann Shannon entropy associated with the MMKE should be
conserved, namely%
\begin{equation}
\frac{\partial S(\rho _{1}^{(N)}(t))}{\partial t}=0,  \label{const H-THM}
\end{equation}%
where $S(\rho _{1}^{(N)}(t))$ denotes again the functional (\ref{B-.S
ENTROPY}).

\subsection{\emph{\ }3.1 - The problem of incompressibility for MMKE}

For definiteness let us first introduce appropriate velocity moments of the $%
1-$body PDF $\rho _{1}^{(N)}(t)$ which are associated with an arbitrary
solution\ of MMKE and the corresponding velocity moments of MMKE. In
particular, introducing the relative velocity $\mathbf{u=v}_{1}-\mathbf{V}%
_{1}(\mathbf{r}_{1},t),$ let us consider the integrals $M_{i}\left[ \rho
_{1}^{(N)}\right] =\int\limits_{U_{1}}d^{3}\mathbf{v}_{1}X_{i}\rho
_{1}^{(N)}(\mathbf{x}_{1},t)$ which correspond to the weight-functions
\begin{equation}
X_{i}=m,m\mathbf{v}_{1},\frac{m}{3}u^{2}  \label{WEIGHT FUNCTIONS}
\end{equation}%
(with the third one identifying the scalar pressure density), namely
\begin{gather}
n_{1}(\mathbf{r}_{1},t)=\int\limits_{U_{1}}d^{3}\mathbf{v}_{1}\rho
_{1}^{(N)}(\mathbf{x}_{1},t),  \label{MOM-0} \\
\rho _{1}(\mathbf{r}_{1},t)=mn_{1}(\mathbf{r}_{1},t),  \label{MOM-1} \\
\mathbf{V}_{1}(\mathbf{r}_{1},t)\equiv \frac{1}{\rho _{1}(\mathbf{r}_{1},t)}%
\int\limits_{U_{1}}d^{3}\mathbf{v}_{1}\mathbf{v}_{1}\rho _{1}^{(N)}(\mathbf{x%
}_{1},t),  \label{MOM-2} \\
p_{1}(\mathbf{r}_{1},t)=\int\limits_{U_{1}}d^{3}\mathbf{v}_{1}\frac{m}{3}%
u^{2}\rho _{1}^{(N)}(\mathbf{x}_{1},t).  \label{MOM-3}
\end{gather}%
Here $\left\{ \rho _{1}(\mathbf{r}_{1},t)\geq 0,\mathbf{V}_{1}(\mathbf{r}%
_{1},t),p_{1}(\mathbf{r}_{1},t)\right\} $ identify respectively the
configuration-space kinetic mass density, velocity and pressure and in
particular it follows that by construction the kinetic pressure is
necessarily strictly positive, \textit{i.e.}, $p_{1}(\mathbf{r}_{1},t)>0$.

Following Refs. \cite{noi1,noi2}, one can show that the vector field $%
\mathbf{F}$ can be (non-uniquely) determined in such a way to satisfy
identically both the isochoricity equation (\ref{INSE-1}) and
incompressibility condition (\ref{incomp0ressibility cond}).\ \ The proof of
the statement follows by suitable prescription of appropriate velocity
moments of MMKE. For this purpose, one first notices that independent of the
choice of the same vector field $\mathbf{F}$, the first velocity moment of
MMKE corresponding to the weight functions $X_{1}=m$ yields simply the
mass-continuity equation, namely
\begin{equation}
\frac{\partial }{\partial t}\rho _{1}+\nabla _{1}\cdot \left( \rho _{1}%
\mathbf{V}_{1}\right) =0.  \label{cont-1}
\end{equation}%
Instead, \ straightforward algebra yields that the velocity moment
associated with $X_{3}=\frac{m}{3}u^{2}$ takes the form:%
\begin{gather}
\frac{\partial p_{1}}{\partial t}+\nabla _{1}\cdot \left( \mathbf{V}%
_{1}p_{1}\right) +\frac{2}{3}\underline{\underline{\mathbf{\Pi }}}:\nabla
_{1}\mathbf{V}+  \notag \\
\nabla _{1}\cdot \mathbf{Q}-\int_{U_{1}}d\mathbf{v}_{1}\frac{2}{3}m\mathbf{%
u\cdot F}\rho _{1}^{(N)}(\mathbf{x}_{1},t)=0,
\end{gather}%
where\ $\underline{\underline{\mathbf{\Pi }}}$ and $\mathbf{Q}$ are
respectively the kinetic tensor pressure

\begin{equation}
\underline{\underline{\mathbf{\Pi }}}=\int_{U_{1}}d\mathbf{v}_{1}m\mathbf{uu}%
\rho _{1}^{(N)}(\mathbf{x}_{1},t)  \label{kinetic tensor pressure}
\end{equation}%
and the relative kinetic energy flux%
\begin{equation}
\mathbf{Q=}\int_{U_{1}}d\mathbf{v}_{1}\mathbf{u}\frac{m}{3}u^{2}\rho
_{1}^{(N)}(\mathbf{x}_{1},t).  \label{kinetic energy flux}
\end{equation}%
Therefore, by suitable prescription of $\mathbf{F}$\textbf{,} one can always
require that%
\begin{gather}
\int_{U_{1}}d\mathbf{v}_{1}\frac{2}{3}m\mathbf{u\cdot F}\rho _{1}^{(N)}(%
\mathbf{x}_{1},t)=  \notag \\
\left[ \frac{2}{3}\underline{\underline{\mathbf{\Pi }}}:\nabla _{1}\mathbf{V}%
+\nabla _{1}\cdot \mathbf{Q}\right] +\frac{Dp}{Dt}.  \label{VINCOLO-1}
\end{gather}%
As a result, the velocity moment associated with $X_{3}=\frac{m}{3}u^{2}$
takes the form%
\begin{equation}
\frac{\partial p_{1}}{\partial t}+\nabla _{1}\cdot \left( \mathbf{V}%
_{1}p_{1}\right) -\frac{Dp_{1}}{Dt}=0,  \label{Fourier-equation}
\end{equation}%
with $\frac{D}{Dt}$ denoting Lagrangian derivative $\frac{D}{Dt}=\frac{%
\partial }{\partial t}+\mathbf{V}_{1}\cdot \nabla _{1}.$ Therefore, provided
$p_{1}>0$ in the open set $\Omega \times I,$ the last equation implies
identically the validity of the isochoricity condition (analogous to Eq.(\ref%
{INSE-1}))
\begin{equation}
\nabla _{1}\cdot \mathbf{V}_{1}=0.  \label{Fourier-equation-2}
\end{equation}%
Then, upon imposing the initial condition
\begin{equation}
\rho _{1}(\mathbf{r}_{1},t_{o})=\rho _{o}>0,
\end{equation}%
with $\rho _{o}$ a non-vanishing constant in the closure $\overline{\Omega }$
of the set $\Omega ,$ the simultaneous validity of Eqs.(\ref{cont-1}) and (%
\ref{Fourier-equation}) requires identically that
\begin{equation}
\rho _{1}(\mathbf{r}_{1},t)=\rho _{o}>0  \label{incompressibility cond-2}
\end{equation}%
must hold in whole the set $\overline{\Omega }\times I,$ thus implying the
validity of the incompressibility condition (\ref{incomp0ressibility cond}).
\

As shown in Refs. NOI1,NOI2, in order to satisfy the constraint equation (%
\ref{Fourier-equation}), this requires to select a vector field $\mathbf{F}$
expressed as a polynomial function of the kinetic relative velocity $\mathbf{%
u=v}_{1}-\mathbf{V}_{1}(\mathbf{r}_{1},t).$ In particular to satisfy the
constraint equation indicated above (see Eq.(\ref{Fourier-equation-2})) it
is sufficient to require that $\mathbf{F}$ is represented by a 3rd degree
polynomial for the type:%
\begin{equation}
\mathbf{F=F}_{0}+\mathbf{F}_{1}+\mathbf{F}_{2}+\mathbf{F}_{3}+\mathbf{F}_{4},
\label{REPRESENTATION}
\end{equation}%
where in particular $\mathbf{F}_{0}$ is independent of $\mathbf{u}$, $%
\mathbf{F}_{1}=\mathbf{u}F_{1},$ is linearly-dependent on $\mathbf{u,}$
while the remaining contributions do not contribute by assumption to the
moment \ref{VINCOLO-1}. Then, in order to satisfy Eq.(\ref{Fourier-equation}%
) it is manifestly sufficient to require that%
\begin{gather}
F_{1}=\frac{1}{2p_{1}(\mathbf{r}_{1},t)}\left[ \frac{2}{3}\underline{%
\underline{\mathbf{\Pi }}}:\nabla _{1}\mathbf{V}_{1}\right.  \notag \\
\left. +\nabla _{1}\cdot \mathbf{Q}\right] +\frac{1}{2}\frac{D\ln p_{1}}{Dt}.
\end{gather}%
In fact this implies
\begin{gather}
\int_{U_{1}}d\mathbf{v}_{1}\frac{2}{3}mu^{2}F_{1}\rho _{1}^{(N)}(\mathbf{x}%
_{1},t)=  \notag \\
\left[ \frac{2}{3}\underline{\underline{\mathbf{\Pi }}}:\nabla _{1}\mathbf{V}%
+\nabla _{1}\cdot \mathbf{Q}\right] +\frac{Dp}{Dt},
\end{gather}%
which fulfills identically (\ref{VINCOLO-1}).

\subsection{\emph{\ }3.2 - The problem of Navier-Stokes equation for MMKE}

Let us now pose the problem of requiring validity of the Navier-Stokes
equation (\ref{INSE-2}). For this purpose let us first evaluate the second
velocity moment of MMKE corresponding to the weight function $m\mathbf{v}%
_{1}.$ Taking into account Eqs.(\ref{Fourier-equation-2}) and (\ref%
{incompressibility cond-2}) elementary algebra yields%
\begin{gather}
\rho _{o}\frac{\partial \mathbf{V}_{1}}{\partial t}+\rho _{o}\mathbf{V}%
_{1}\cdot \nabla _{1}\mathbf{V}_{1}+  \notag \\
\nabla _{1}\cdot \underline{\underline{\mathbf{\Pi }}}-\int_{U_{1}}d\mathbf{v%
}_{1}\mathbf{F}\rho _{1}^{(N)}(\mathbf{x}_{1},t)=0  \label{moment-3}
\end{gather}%
Finally let us require, in particular, that $\mathbf{F}$ is prescribed so
that
\begin{gather}
\int_{U_{1}}d\mathbf{v}_{1}\mathbf{F}\rho _{1}^{(N)}(\mathbf{x}%
_{1},t)=\nabla _{1}\cdot \underline{\underline{\mathbf{\Pi }}}-  \notag \\
\nabla _{1}p_{1}+\mathbf{f}_{1}+\mu \nabla _{1}^{2}\mathbf{V}_{1}.
\label{VINCOILO-2}
\end{gather}%
As a consequence it follows that Eq.(\ref{moment-3}) takes a form analogous
to the Navier-Stokes equation (\ref{INSE-2}), namely%
\begin{gather}
\rho _{o}\frac{\partial \mathbf{V}_{1}}{\partial t}+\rho _{o}\mathbf{V}%
_{1}\cdot \nabla _{1}\mathbf{V}_{1}+  \notag \\
\nabla _{1}p_{1}-\mathbf{f}_{1}-\mu \nabla _{1}^{2}\mathbf{V}_{1}=0.
\label{KINETIC NAVIER-STOKES}
\end{gather}%
To recover, however in a proper sense also the proper values of the
Navier-Stokes fluid fields $\{\rho ,\mathbf{V},p\}$ a suitable mapping must
be introduced to relate them to the kinetic moments $\{\rho _{1},\mathbf{V}%
_{1},p_{1}\}$ indicated above. More precisely, such a relationship is
realized by the \emph{kinetic correspondence principle:}%
\begin{equation}
\left\{
\begin{array}{c}
\rho (\mathbf{r}_{1},t)=\rho _{1}(\mathbf{r}_{1},t)=\rho _{o}, \\
\mathbf{V}(\mathbf{r}_{1},t)=\mathbf{V}_{1}(\mathbf{r}_{1},t)+\mathbf{V}_{2}(%
\mathbf{r}_{1}), \\
p(\mathbf{r}_{1},t)=p_{1}(\mathbf{r}_{1},t)+p_{2}(\mathbf{r}_{1}),%
\end{array}%
\right.  \label{CORRESPONDENCE PRINCUJPLE}
\end{equation}%
in which the fields $\left\{ \rho _{o},\mathbf{V}_{2}(\mathbf{r}_{1}),p_{2}(%
\mathbf{r}_{1})\right\} $ identify here an - in principle arbitrary -
particular stationary solution of INSE, namely such that identically in $%
\Omega \times I:$%
\begin{equation}
\left\{
\begin{array}{c}
\nabla _{1}\cdot \mathbf{V}_{2}(\mathbf{r}_{1})=0, \\
\rho _{o}\mathbf{V}_{2}(\mathbf{r}_{1})\cdot \nabla _{1}\mathbf{V}_{2}(%
\mathbf{r}_{1})+\nabla _{1}p_{2}-\mathbf{f}_{2}-\mu \nabla _{1}^{2}\mathbf{V}%
_{2}=0.%
\end{array}%
\right.  \label{STATIONARY INSE}
\end{equation}%
Therefore, from the initial conditions holding for the Navier-Stokes fluid
fields (\ref{INITIAL COND}), it follows that the kinetic moments $\{\rho
_{1},\mathbf{V}_{1},p_{1}\}$ must fulfill the initial conditions%
\begin{equation}
\left\{
\begin{array}{c}
\rho _{1}(\mathbf{r}_{1},t_{o})=\rho _{o}, \\
\mathbf{V}_{1}(\mathbf{r}_{1},t_{o})=\mathbf{V}_{o}(\mathbf{r}_{1})-\mathbf{V%
}_{2}(\mathbf{r}_{1}), \\
p_{1}(\mathbf{r}_{1},t_{o})=p_{o}(\mathbf{r}_{1})-p_{2}(\mathbf{r}_{1}).%
\end{array}%
\right.  \label{INITIAL COND-KINETIC}
\end{equation}%
Notice that in Eqs.(\ref{STATIONARY INSE}) $p_{2}(\mathbf{r}_{1})$ is
actually defined up to an arbitrary constant. Therefore one can always
require the initial kinetic pressure $p_{1}(\mathbf{r}_{1},t_{o})$ to be
strictly positive, \textit{i.e.},
\begin{equation}
p_{1}(\mathbf{r}_{1},t_{o})=p_{o}(\mathbf{r}_{1})-p_{2}(\mathbf{r}_{1})>0.
\end{equation}%
Based on the representation (\ref{REPRESENTATION}), for the fulfillment of
the constraint equation (\ref{VINCOILO-2}) it is manifestly sufficient to
require%
\begin{equation}
\mathbf{F}_{0}=\frac{1}{\rho _{o}}\left[ \nabla _{1}\cdot \underline{%
\underline{\mathbf{\Pi }}}-\nabla _{1}p+\mu \nabla _{1}^{2}\mathbf{V}_{1}%
\right] ,
\end{equation}%
while also requiring that%
\begin{equation}
\int_{U_{1}}d\mathbf{v}_{1}\mathbf{F}\rho _{1}^{(N)}(\mathbf{x}%
_{1},t)=\int_{U_{1}}d\mathbf{v}_{1}\mathbf{F}_{0}\rho _{1}^{(N)}(\mathbf{x}%
_{1},t),
\end{equation}%
\textit{i.e.}, the remaining terms ($\mathbf{F}_{2},\mathbf{F}_{3}$ and $%
\mathbf{F}_{4}$) in the representation (\ref{REPRESENTATION}) do not
contribute, so that identically
\begin{equation*}
\int_{U_{1}}d\mathbf{v}_{1}\mathbf{F}_{0}\rho _{1}^{(N)}(\mathbf{x}%
_{1},t)=\nabla _{1}\cdot \underline{\underline{\mathbf{\Pi }}}-\nabla
_{1}p-\mu \nabla _{1}^{2}\mathbf{V.}
\end{equation*}

\subsection{3.3 - The problem of entropy conservation for MMKE}

Let us now pose the problem of determining a possible non-unique realization
of the remaining terms in the polynomial expansion (\ref{REPRESENTATION}) in
such a way to fulfill the constant entropy theorem (\ref{const H-THM}). For
definiteness we consider here only the additional contributions arising from
$\mathbf{F}_{2},\mathbf{F}_{3}$ and $\mathbf{F}_{4},$ more precisely
requiring that they are of the form see
\begin{eqnarray}
\mathbf{F}_{2} &=&\mathbf{F}_{2}(\mathbf{r}_{1},t),  \notag \\
\mathbf{F}_{3} &=&\frac{mu^{2}}{3p_{1}}F_{3}(\mathbf{r}_{1},t),  \notag \\
\mathbf{F}_{4} &=&\frac{m\mathbf{u}}{3p_{1}}F_{4}(\mathbf{r}_{1},t).
\label{REPRESENTATION-2}
\end{eqnarray}%
Here the real vector/scalar fields $\mathbf{F}_{2}(\mathbf{r}_{1},t),$ $%
F_{3}(\mathbf{r}_{1},t)$ and $F_{4}(\mathbf{r}_{1},t)$ still remain in
principle completely arbitrary. Starting point is provided by the
theoretical results established Refs. \cite{noi3,noi4}. These warrant the
conservation laws of the Master collision operator, and in particular also
the validity of a constant $H-$theorem analogous to Eq. (\ref{const H-THM})
for the Master kinetic equation (\textit{i.e.}, when the mean-field
interaction $\mathbf{F}$ vanishes identically). However, the same $H-$%
theorem is generally now warranted in case of the MMKE. This follows from
direct evaluation of the phase-space moment of MMKE determined in terms of
the weight-function $X=\ln \rho _{1}^{(N)}(\mathbf{x}_{1},t),$ yielding the
Boltzmann-Shannon entropy (\ref{B-.S ENTROPY}). In fact, elementary algebra
yields in the case of the MMKE:%
\begin{equation}
\frac{\partial S(\rho _{1}^{(N)}(t))}{\partial t}=K_{1}(\rho
_{1}^{(N)}(t))+K_{F}(\rho _{1}^{(N)}(t)),
\end{equation}%
where the entropy production rate $K_{1}(\rho _{1}^{(N)}(t))$ is defined by
Eq.(\ref{K_1}) and hence vanishes identically in the case of the Master
collision operator \cite{noi3,noi4} (see Eq.(\ref{IDENRITTY FOR K_1}) in
subsection 2.2); furthermore $K_{F}(\rho _{1}^{(N)}(t))$ is given by
\begin{equation}
K_{F}(\rho _{1}^{(N)}(t))=-\int_{\Gamma _{1}}d\mathbf{x}_{1}\rho _{1}^{(N)}(%
\mathbf{x}_{1},t)\frac{\partial }{\partial \mathbf{v}_{1}}\cdot \mathbf{F}
\label{K_F}
\end{equation}%
and hence, unless a further specific constraint is placed on the choice of
the vector field $\mathbf{F,}$ it is generally non-vanishing.\ Therefore the
question is whether the remaining contributions $\mathbf{F}_{2},\mathbf{F}%
_{3}$ and $\mathbf{F}_{4}$ in the polynomial representation (\ref%
{REPRESENTATION}) of the mean-field interaction $\mathbf{F}$ can be
prescribed in such a way to fulfill the further constraint condition%
\begin{equation}
\int_{\Gamma _{1}}d\mathbf{x}_{1}\rho _{1}^{(N)}(\mathbf{x}_{1},t)\frac{%
\partial }{\partial \mathbf{v}_{1}}\cdot \mathbf{F}=0  \label{VINCOLO-3}
\end{equation}%
(\emph{entropic constraint}). However, one should add for consistency also
the requirement that the same terms ($\mathbf{F}_{2},\mathbf{F}_{3}$ and $%
\mathbf{F}_{4}$) should leave unaffected the two previous constraint
conditions (\ref{VINCOLO-1}) and (\ref{VINCOILO-2}). In view of the
representation (\ref{REPRESENTATION}) and Eqs. (\ref{REPRESENTATION-2}) it
follows in particular:
\begin{gather}
\frac{\partial }{\partial \mathbf{v}_{1}}\cdot \mathbf{F}\mathbf{=}\frac{3}{%
2p_{1}}\left[ \nabla _{1}\cdot \mathbf{Q+}\frac{2}{3}\nabla _{1}\mathbf{V:}%
\underline{\underline{\mathbf{\Pi }}}\right] +  \notag \\
3\frac{D\ln p_{1}}{Dt}+\frac{m}{p}F_{4}+\left[ \frac{mu^{2}}{p_{1}}+\frac{%
2mu^{2}}{3p_{1}}\right] F_{3}.  \label{divergenza}
\end{gather}%
This means that the vector and scalar fields $\mathbf{F}_{2},F_{4}$ and $%
F_{3}$ should be prescribed in such a way to fulfill identically in $\Omega
\times I$ the following velocity-moment constraint equations
\begin{eqnarray}
\int_{U_{1}}d\mathbf{v}_{1}\left[ \mathbf{F}_{2}+\mathbf{u}\frac{mu^{2}}{%
3p_{1}}F_{3}\right] \rho _{1}^{(N)}(\mathbf{x}_{1},t) &=&\mathbf{0,} \\
\int_{U_{1}}d\mathbf{v}_{1}\mathbf{u\cdot }\left[ \frac{m\mathbf{u}}{3p_{1}}%
F_{4}+\mathbf{u}\frac{mu^{2}}{3p_{1}}F_{3}\right] \rho _{1}^{(N)}(\mathbf{x}%
_{1},t) &=&0.
\end{eqnarray}%
Then a sufficient condition in order to satisfy the entropic constraint
indicated above is manifestly to require validity in the whole set $\Omega
\times I$ of the following integral identity
\begin{equation}
\int_{U_{1}}d\mathbf{v}_{1}\rho _{1}^{(N)}(\mathbf{x}_{1},t)\frac{\partial }{%
\partial \mathbf{v}_{1}}\cdot \mathbf{F}=0.  \label{VINCOLO-3BIS}
\end{equation}%
Evaluation of the velocity integrals indicated above yields therefore the
following constraint equations for the still undetermined vector and scalar
fields $\mathbf{F}_{2},F_{3}$ and $F_{4}$:%
\begin{gather}
n_{1}(\mathbf{r}_{1},t)\mathbf{F}_{2}+\frac{1}{p_{1}}\mathbf{Q}F_{3}=0, \\
F_{4}+\frac{1}{p_{1}}WF_{3}=0, \\
\frac{3n_{1}(\mathbf{r}_{1},t)}{2p_{1}}\left[ \nabla _{1}\cdot \mathbf{Q+}%
\frac{2}{3}\nabla _{1}\mathbf{V:}\underline{\underline{\mathbf{\Pi }}}\right]
+  \notag \\
3n_{1}(\mathbf{r}_{1},t)\frac{D\ln p_{1}}{Dt}+\frac{mn_{1}(\mathbf{r}_{1},t)%
}{p}F_{4}+5F_{3}=0,
\end{gather}%
where
\begin{equation}
W(\mathbf{r}_{1},t)=\int_{U_{1}}d\mathbf{v}_{1}\frac{mu^{4}}{3}\rho
_{1}^{(N)}(\mathbf{x}_{1},t).
\end{equation}%
The solution of the first equation gives for $\mathbf{F}_{2}$ a unique
prescription in terms of the scalar field $F_{3},$ namely
\begin{equation}
\mathbf{F}_{2}=-\frac{\mathbf{Q}}{n_{1}(\mathbf{r}_{1},t)}F_{3}.
\label{ROOT-1a}
\end{equation}%
Then, provided
\begin{equation}
\Delta \equiv 5-\frac{mn_{1}(\mathbf{r}_{1},t)}{p_{1}^{2}}W>0,
\label{INEQ-4a}
\end{equation}%
the functions $F_{3}$ and $F_{4}$ are given respectively by

\begin{eqnarray}
F_{4} &=&\frac{1}{p_{1}}WF_{3},  \label{ROOT-2} \\
F_{3} &=&\frac{S_{3}(\mathbf{r}_{1},t)}{5-\frac{mn_{1}(\mathbf{r}_{1},t)}{%
p_{1}^{2}}W},  \label{ROOT-3}
\end{eqnarray}%
with
\begin{align}
S_{3}(\mathbf{r}_{1},t)& \equiv \frac{3n_{1}(\mathbf{r}_{1},t)}{2p_{1}}\left[
\nabla _{1}\cdot \mathbf{Q+}\frac{2}{3}\nabla _{1}\mathbf{V:}\underline{%
\underline{\mathbf{\Pi }}}\right] +  \notag \\
& 3n_{1}(\mathbf{r}_{1},t)\frac{D\ln p_{1}}{Dt}.
\end{align}%
Regarding the validity of the requirement (\ref{INEQ-4a}) one notices that,
thanks to Schwartz's inequality, necessarily
\begin{equation}
\frac{mn_{1}(\mathbf{r}_{1},t)}{p_{1}^{2}}W\leq 3
\end{equation}%
must hold, thus implying in turn\ the validity of the inequality (\ref%
{INEQ-4a})$.$ Therefore, Eqs.(\ref{ROOT-1a}),(\ref{ROOT-2}) and (\ref{ROOT-3}%
) realize a unique solution to the constraint condition (\ref{VINCOLO-3})
which warrants the validity of the constant H-theorem (\ref{const H-THM}).

\bigskip

\section{3.4 - Implications}

Let us briefly summarize the implications of subsections 3.1-3.3. Based on
the kinetic correspondence principle (\ref{CORRESPONDENCE PRINCUJPLE}) and
the moment equations of MMKE determined for the weight-functions (\ref%
{WEIGHT FUNCTIONS}) it follows that the isochoricity and incompressibility
conditions are fulfilled respectively by the kinetic velocity moment $%
\mathbf{V}_{1}(\mathbf{r}_{1},t)$ (\ref{MOM-2}) and the corresponding
kinetic/fluid mass density moment $\rho _{o}$ (\ref{MOM-1}) given
respectively by Eqs. (\ref{Fourier-equation-2}) and (\ref{incomp0ressibility
cond}) (subsection 3.1). As a consequence, the Navier-Stokes equation (\ref%
{KINETIC NAVIER-STOKES}) is satisfied by the same kinetic velocity moment $%
\mathbf{V}_{1}(\mathbf{r}_{1},t)$ and the kinetic scalar pressure $p_{1}$ (%
\ref{MOM-3}), so that the kinetic moments $\{\rho _{1},\mathbf{V}%
_{1},p_{1}\} $ provide a particular solution of INSE (see Eqs.(\ref{INSE-1})
and (\ref{INSE-2})). However, in order to satisfy the initial conditions (%
\ref{INITIAL COND}, and in particular the corresponding ones holding for the
kinetic moments one must generally identify the Navier-Stokes fluid fields $%
\{\rho ,\mathbf{V},p\}$ in terms of the kinetic moments $\{\rho _{1},\mathbf{%
V}_{1},p_{1}\}$ determined above only via the kinetic correspondence
principle (\ref{CORRESPONDENCE PRINCUJPLE}). Therefore, Eqs. (\ref%
{CORRESPONDENCE PRINCUJPLE}) yield the general solution of INSE (\ref{INSE-1}%
) and (\ref{INSE-2}) holding globally in $\Omega \times I$ and subject to
the initial and boundary conditions provided by Eqs. (\ref{INITIAL
COND-KINETIC}) and (\ref{BOUNDARY COND}). Finally, by suitably prescribing
the mean-field interaction $\mathbf{F}$ (see Eqs. (\ref{REPRESENTATION}) and
(\ref{REPRESENTATION-2})), the entropic constraint (\ref{VINCOLO-3}) is
identically fulfilled thanks to the local condition (\ref{VINCOLO-3BIS}).
This means that by construction the Boltzmann-Shannon entropy associated to
an arbitrary solution of MMKE is conserved.

\bigskip

\section{4 - Example applications of MMKE: the plane Couette and the
Poiseuille flows}

Let us now test the validity of the statistical description of the INSE
problem based on MMKE. Indeed, an interesting issue is whether one can
recover in this way well-known particular possible realizations of
incompressible Navier-Stokes fluids. Here two examples are considered which
concern respectively:

\begin{enumerate}
\item \emph{the plane Couette flow}, represented by the solution%
\begin{equation}
\left\{
\begin{array}{c}
\rho (\mathbf{r}_{1},t)=\rho _{o}, \\
p_{2}(\mathbf{r}_{1})=p_{o}, \\
\mathbf{V}_{2}\mathbf{(\mathbf{r}_{1}})\mathbf{=}V_{o}y\widehat{\mathbf{x}}.%
\end{array}%
\right.  \label{COUETTE}
\end{equation}%
with $\mathbf{\mathbf{r}}_{1}\equiv \left( x,y,z\right) $ and $p_{o}>0$ a
constant. The corresponding INSE are realized by the equations%
\begin{equation}
\left\{
\begin{array}{c}
V_{o}\nabla _{1}\cdot y\widehat{\mathbf{x}}=0, \\
\rho _{o}V_{o}^{2}y\widehat{\mathbf{x}}\cdot \nabla _{1}y\widehat{\mathbf{x}}%
-\mu V_{o}\nabla _{1}^{2}y\widehat{\mathbf{x}}=0.%
\end{array}%
\right.  \label{COUETTE-1}
\end{equation}

\item \emph{the plane Poiseuille flow}, represented by the solution%
\begin{equation}
\left\{
\begin{array}{c}
\rho (\mathbf{r}_{1},t)=\rho _{o}, \\
p_{2}(\mathbf{\mathbf{r}_{1}})=p_{o}+\beta x, \\
\mathbf{V}_{2}(\mathbf{\mathbf{\mathbf{r}_{1}}})\mathbf{=}V_{o}\left(
1-y^{2}\right) \widehat{\mathbf{x}}%
\end{array}%
\right.  \label{POISELLE}
\end{equation}

with $\beta >0$ a constant. Notice, however, that the pressure indicated in
Eq.(\ref{POISELLE}) is equivalent to the one occurring due to a constant (%
\textit{i.e.}, equilibrium) gravitational force acting along the $x-$%
direction. Therefore the corresponding set of INSE can be identified with%
\begin{equation}
\left\{
\begin{array}{c}
V_{o}\nabla _{1}\cdot y\widehat{\mathbf{x}}=0, \\
\rho _{o}V_{o}^{2}y\left( 1-y^{2}\right) \widehat{\mathbf{x}}\cdot \nabla
_{1}\left( 1-y^{2}\right) \widehat{\mathbf{x}}- \\
\mu V_{o}\nabla _{1}^{2}\left( 1-y^{2}\right) \widehat{\mathbf{x}}=0.%
\end{array}%
\right.  \label{POISEUILLE-1}
\end{equation}
\end{enumerate}

Let us examine separately the two cases.

The first one is of immediate realization. In fact the fluid fields (\ref%
{COUETTE}) correspond manifestly to a particular stationary solution $%
\left\{ \rho _{o},\mathbf{V}_{2}(\mathbf{r}_{1}),p_{2}(\mathbf{r}%
_{1})\right\} $ of INSE (see Eqs. (\ref{STATIONARY INSE})) in which the
fluid pressure is constant and positive while the fluid domain is unbounded
and two dimensional. The second one is analogous. The only difference arises
because now the pressure can become negative and is unbounded too. These
features, however - namely the occurrence of an unbounded fluid domain $%
\Omega $ and possibly also the appearance of an unbounded stationary fluid
pressure - do not pose any restriction on the validity of the kinetic theory
here developed. Therefore, in both cases the kinetic correspondence
principle (\ref{CORRESPONDENCE PRINCUJPLE}) provides a representation of the
general solution of INSE which admits either Eqs.(\ref{COUETTE}) or (\ref%
{STATIONARY INSE}) as particular stationary solutions.

We remark that in both cases the fluid domain $\Omega $ is realized in
principle by a $2-$dimensional infinite strip between two parallel
co-oriented straight lines lying at distance $H$, $y$ being the Cartesian
coordinates orthogonal to both ones and $x$ the Cartesian coordinate along
the same lines. Notice, however, that Eqs. (\ref{COUETTE-1}) and (\ref%
{POISEUILLE-1}) can be equivalently realized by replacing $\Omega $ with a
bounded domain $\Omega _{p}$ identified with a rectangle of finite width $L$
along the $x-$direction and introducing for the fluid velocity $\mathbf{V}(%
\mathbf{\mathbf{\mathbf{r}_{1}}},t)$ periodic boundary conditions at the two
boundaries located respectively at $x=0$ and $x=L,$ namely letting%
\begin{equation}
\mathbf{V}(x=0,y,t)\mathbf{=V}(x=L,y,t).
\end{equation}%
Under these conditions, therefore, the fluid domain $\Omega \equiv \Omega
_{p}$ is bounded so that integration of the Navier-Stokes equation (see (\ref%
{INSE-2})) delivers in a straightforward way the time-evolution of the
kinetic energy of the fluid, namely%
\begin{gather}
\frac{\partial }{\partial t}\int_{\Omega p}d^{3}\mathbf{r}_{1}\rho
_{o}^{2}V^{2}(\mathbf{\mathbf{\mathbf{r}_{1}}},t)=  \notag \\
-\mu \int_{\Omega p}d^{3}\mathbf{r}_{1}\nabla _{1}\mathbf{V}(\mathbf{\mathbf{%
\mathbf{r}_{1}}},t)\mathbf{:}\nabla _{1}\mathbf{V}(\mathbf{\mathbf{\mathbf{r}%
_{1}}},t)<0.
\end{gather}%
This implies necessarily that $\mathbf{V}(\mathbf{\mathbf{\mathbf{r}_{1}}}%
,t) $ must decay asymptotically to a stationary solution, namely%
\begin{equation}
\lim_{t\rightarrow +\infty }\mathbf{V}(\mathbf{\mathbf{\mathbf{r}_{1}}},t)=%
\mathbf{V}_{2}\mathbf{(\mathbf{r}_{1}}),  \label{DECAY-PHE-1}
\end{equation}%
with $\mathbf{V}_{2}\mathbf{(\mathbf{r}_{1}})$ corresponding either to Eqs. (%
\ref{COUETTE}) or (\ref{POISELLE}), In a similar way, thanks to Eq.(\ref%
{POISSON}), one expects also the decay of the related fluid pressure $p(%
\mathbf{\mathbf{\mathbf{r}_{1}}},t),$ namely%
\begin{equation}
\lim_{t\rightarrow +\infty }p(\mathbf{\mathbf{\mathbf{r}_{1}}},t)=p_{2}%
\mathbf{(\mathbf{r}_{1}}),  \label{DECAY-PHE-2}
\end{equation}%
to occur.

The two example cases indicated above have, therefore, an important physical
implication. This is precisely realized by the \emph{fluid decay conditions}
(\ref{DECAY-PHE-1})-(\ref{DECAY-PHE-2}), \textit{i.e.}, from the fact that
both the stationary planar Couette and Poiseuille flows actually coincide
with the final equilibrium states of the fluid. In other words in both cases
they are the result of the decay of a suitable non-stationary Navier-Stokes
fluids $\left\{ \mathbf{V}(\mathbf{\mathbf{\mathbf{r}_{1}}},t),p(\mathbf{%
\mathbf{\mathbf{r}_{1}}},t)\right\} $ which, in turn, both correspond in
principle to arbitrary initial conditions (\ref{INITIAL COND}).

\section{5 - Decay to kinetic equilibrium for MMKE}

The interesting issue which arises is whether the fluid decay phenomenon
indicated above may correspond or not to the occurrence of a \emph{global
decay to kinetic equilibrium} (DKE) for the relevant $1-$body kinetic PDF,
\textit{i.e.}, occurring globally in the phase space $\Gamma _{1}.$ More
precisely, this means that time-dependent $1-$body kinetic PDF solution of
MMKE should decay uniformly to a stationary solution of MMKE, in the sense
that provided $\rho _{1}^{(N)}(\mathbf{x}_{1},t)$ is suitably smooth in the
extended phase-space $\Gamma _{1}\times I$ it should occur uniformly in $%
\Gamma _{1}$ that:
\begin{equation}
\lim_{t\rightarrow +\infty }\rho _{1}^{(N)}(\mathbf{x}_{1},t)=\rho
_{M}^{(N)}(\mathbf{v}_{1}),  \label{KINETIC DECAY LIMIT}
\end{equation}%
\textit{i.e.}, in other words $\rho _{1}^{(N)}(\mathbf{x}_{1},t)$ should
decay asymptotically to a prescribed stationary $1-$body kinetic PDF. As
shown in the Appendix $\rho _{M}^{(N)}(\mathbf{v}_{1})$ this means that it
should necessarily coincide with a \emph{spatially uniform kinetic
equilibrium,} which is realized by \emph{a spatially-uniform local
Maxwellian PDF}%
\begin{equation}
\rho _{M}^{(N)}(\mathbf{v}_{1})=n_{o}\frac{1}{\left( \pi v_{th}\right) ^{3/2}%
}\exp \left\{ -\frac{\mathbf{v}_{1}^{2}}{v_{th}^{2}}\right\} .
\label{MAXWELLIAN-1}
\end{equation}%
Here the notation is standard. Thus
\begin{equation}
v_{th}^{2}=2p_{1o}/mn_{o}
\end{equation}%
is the thermal velocity, $n_{o}$ the constant configuration-space
probability density, $m$ the mass of a hard sphere and $p_{1o}>0$ an
arbitrary non-vanishing constant kinetic pressure.

As shown in the Appendix, the validity of the kinetic decay limit (\ref%
{KINETIC DECAY LIMIT}) for MMKE arises because, in validity of the fluid
decay conditions (\ref{DECAY-PHE-1})-(\ref{DECAY-PHE-2}), the only
admissible kinetic equilibrium is provided by Eq. (\ref{MAXWELLIAN-1}).

This conclusion appears, in our view, interesting and in some respects also
surprising.

The notable aspect lies in the physical mechanism at the basis of the
global-DKE phenomenology. In fact it is manifest that, at least for case of
the MMKE considered here, the phenomenon of DKE occurs specifically as a
consequence of collisions occurring within the particles of the hard-sphere$%
\ N-$body system here described by means of the Master collision operator.
This warrants that the entropy production rate associated with by the same
collision operator vanishes identically (see subsection 2.2). Nevertheless
as shown above the Boltzmann-Shannon entropy remains constant by
construction also for MMKE, namely when the mean-field force $\mathbf{F\ }$%
is introduced and a proper prescription is adopted for the same vector field.

This conclusion is potentially in conflict with the customary interpretation
of Boltzmann kinetic theory. In fact, in accordance with Boltzmann H-theorem
the physical origin of DKE is usually ascribed to the macroscopic
irreversibility property of the Boltzmann-Shannon entropy functional
together with the requirement that the same quantity should remain finite
also in the time-asymptotic limit $t\rightarrow +\infty $.

Nevertheless, according to Boltzmann's own original interpretation,
Boltzmann equation and Boltzmann H-theorem are only supposed to hold when
both the dilute gas asymptotic ordering and the continuum Boltzmann-Grad
limit apply, while for according to Boltzmann's own conjecture finite-size
hard sphere the Boltzmann-Shannon entropy "\textit{should tend to a constant}%
" . These viewpoints are included in Boltzmann's replies to Zermelo
(1896-1897 \cite{Boltzmann1896,Boltzmann1896c,Boltzmann1897}). \ In other
words, the same equations cannot hold when the hard-spheres are considered
finite-sized as in the present case. Indeed, in striking departure from
Boltzmann kinetic theory, MMKE holds for an arbitrary finite Boltzmann-Sinai
CDS, i.e., which both the number of particles $N$ and their diameter $\sigma
$ are considered as finite. Nevertheless, the phenomenon of global DKE
occurs also in the present case independent of the time-behavior of the
Boltzmann-Shannon entropy\emph{.}

The key differences arising between the two theories, \textit{i.e.}, the one
based on the Boltzmann equation and the other discussed here based on the
modified Master kinetic equation, are of course related to the different and
peculiar intrinsic properties of the Boltzmann and Master collision
operators. In particular, as discussed elsewhere (see Refs.\cite%
{noi3,noi6,noi7}), precisely because the Boltzmann equation is only an
asymptotic approximation of the Master kinetic equation explains why a loss
of information occurs in Boltzmann kinetic theory and consequently the
related Boltzmann-Shannon entropy is not conserved.

Nevertheless, as recalled in Section 2, the Boltzmann-Shannon entropy for
the Master kinetic equation is exactly conserved due to the symmetry
properties of the Master collision operator (see subsection 2.2). Therefore
its behavior is manifestly unrelated and independent from the occurrence of
the phenomenon of DKE here pointed out. The present investigation shows,
notwithstanding, that a macroscopic irreversibility property which is
realized by DKE actually occurs. \ As indicated above this can be explained
at a fundamental level, \textit{i.e.}, based specifically on the collision
processes described by the Master collision operator.

\section{6 - Conclusions}

In this paper the problem has been addressed of identifying a possible
microscopic statistical description for the incompressible Navier-Stokes
equations. The statistical approach has been based on the axiomatic
statistical theory of the Boltzmann-Sinai classical dynamical system
recently developed \cite{noi1,noi2,noi3,noi4,noi5,noi6,noi7}.

The theory presented here departs significantly, in several respects, from
previous literature and notably from approaches based on the Boltzmann
kinetic theory. The main differences actually arise because of the
non-asymptotic character of the new theory, \textit{i.e.}, the fact that it
applies to arbitrary dense or rarefied systems for which the finite number
and size of the constituent particles is accounted for \cite{noi3}. In this
paper basic consequences of the new theory have been investigated which
concern the statistical treatment of an incompressible Navier-Stokes fluid
based on the adoption of the Master kinetic equation and the introduction of
a suitable mean-field interaction acting on a system of finite-sized smooth
hard spheres. The phenomenon of decay to global kinetic equilibrium (DKE)
has been pointed out. Remarkably, despite the fact that the related
Boltzmann-Shannon entropy remains preserved, DKE occurs specifically because
of the effect of collisions which are taken into account in the Master
collision operator.

The present results are believed to be crucial both from the theoretical
viewpoint and for applications of the "\textit{ab initio}" statistical
theory, \textit{i.e.}, the Master kinetic equation.\ Indeed, regarding
possible challenging future developments of the theory one should mention
among others the following examples of possible (and mutually-related)
directions worth to be explored:

\begin{itemize}
\item One is the possible ubiquitous occurrence of the DKE phenomenon for
the Master kinetic equation, i.e., even in the absence of the mean-field
interaction $\mathbf{F}$ introduced here.

\item The second is related to the investigation of the time-asymptotic
properties of the same kinetic equation, for which the present paper may
represent a useful basis.

\item The third goal refers to the possible extension of the theory to
mixtures formed by hard spheres of different type, i.e., with different
masses, diameters and undergoing either elastic or inelastic collisions.

\item The fourth one concerns the investigation of hydrodynamic regimes for
which a key prerequisite is provided by the DKE theory here established.
\end{itemize}

\section{Acknowledgments}

Work developed within the research projects of the Albert Einstein Center
for Gravitation and Astrophysics, Czech Science Foundation No. 14-37086G
(M.T.). The initial motivation of the work is related to an earlier
collaboration with the Danieli Co. R\&D Department (Buttrio, Udine, Italy).
The authors are grateful to the International Center for Theoretical Physics
(Miramare, Trieste, Italy) for the hospitality during the preparation of the
manuscript. One of us (M.T.) wishes to thank Dr. David Mukamel (Weizmann
Institute of Science, Rehovot, Israel) for stimulating discussions and
drawing attention to the problem of DKE.

\section{Appendix: Conditions of validity of kinetic equilibrium for MMKE}

In this appendix we pose the problem of establishing necessary conditions
for the existence of stationary solutions, or so-called\emph{\ kinetic
equilibrium solutions}, of MMKE, \textit{i.e.}, such that%
\begin{equation}
\frac{\partial }{\partial t}\rho _{1}^{(N)}(\mathbf{x}_{1},t)\equiv 0
\end{equation}%
namely

\begin{eqnarray}
\mathbf{v}_{1}\cdot \nabla _{1}\rho _{1}^{(N)}(\mathbf{x}_{1},t)+\frac{%
\partial }{\partial \mathbf{v}_{1}}\cdot \left( \mathbf{F}\rho _{1}^{(N)}(%
\mathbf{x}_{1},t)\right) &=&0, \\
\mathcal{C}_{1}\left( \rho _{1}^{(N)}|\rho _{1}^{(N)}\right) &=&0.
\end{eqnarray}%
Let us introduce in addition, consistent with the fluid decay conditions (%
\ref{DECAY-PHE-1})-(\ref{DECAY-PHE-2}), the requirements
\begin{eqnarray}
\lim_{t\rightarrow +\infty }n_{1}(\mathbf{r}_{1},t)
&=&\int\limits_{U_{1}}d^{3}\mathbf{v}_{1}\lim_{t\rightarrow +\infty }\rho
_{1}^{(N)}(\mathbf{x}_{1},t)=n_{o}>0,  \label{C-1a} \\
\lim_{t\rightarrow +\infty }\mathbf{V}_{1}(\mathbf{r}_{1},t) &\equiv &\frac{1%
}{n_{o}}\int\limits_{U_{1}}d^{3}\mathbf{v}_{1}\mathbf{v}_{1}\lim_{t%
\rightarrow +\infty }\rho _{1}^{(N)}(\mathbf{x}_{1},t)\equiv 0.  \label{C-2a}
\end{eqnarray}%
Then it follows that necessarily the $1-$body PDF $\rho _{1}^{(N)}$ must
coincide with $\rho _{1}^{(N)}\equiv \rho _{M}^{(N)}(\mathbf{v}_{1}),$
namely the kinetic equilibrium local Maxwellian PDF (\ref{MAXWELLIAN-1}).
Conversely, let us assume that the fluid velocity moments associated with
the MMKE are identified with the stationary equations given by Eqs. (\ref%
{STATIONARY INSE}), namely are such that, besides Eqs.(\ref{C-1a}) and (\ref%
{C-2a}), they satisfy identically in $\Omega \times I$ also the constraint$:$

\begin{equation}
\lim_{t\rightarrow +\infty }p_{1}(\mathbf{r}_{1},t)==p_{1o}.  \label{C-3a}
\end{equation}%
where $p_{1}(\mathbf{r}_{1},t)=\int\limits_{U_{1}}d^{3}\mathbf{v}_{1}\frac{m%
}{3}\left( \mathbf{v}_{1}-\mathbf{V}_{1}(\mathbf{r}_{1},t)\right) ^{2}\rho
_{1}^{(N)}(\mathbf{x}_{1},t).$ The issue if whether $\lim_{t\rightarrow
+\infty }\rho _{1}^{(N)}(\mathbf{x}_{1},t)$ must necessarily coincide or not
with the Maxwellian PDF given above (\ref{MAXWELLIAN-1}), or in other words
whether the $1-$body PDF $\rho _{1}^{(N)}$ can still be a non-stationary
solution of the MMKE once the constraints Eqs. (\ref{C-1a}),(\ref{C-2a}) and
(\ref{C-3a}) are placed on the same PDF. For definiteness let us consider
the particular case in which for $t\rightarrow +\infty $%
\begin{equation}
\rho _{1}^{(N)}(\mathbf{x}_{1},t)=\rho _{M}^{(N)}(\mathbf{v}_{1})+\delta
\rho _{1}^{(N)}(\mathbf{x}_{1},t),  \label{POSITION-1}
\end{equation}%
where $\delta \rho _{1}^{(N)}(\mathbf{x}_{1},t)$ being an infinitesimal
perturbation of order $\varepsilon ,$ $\varepsilon $ denoting a prescribed
real infinitesimal. Without loss of generality the latter can always be
represented in terms of a polynomial expansion of the type%
\begin{eqnarray}
\delta \rho _{1}^{(N)}(\mathbf{x}_{1},t) &=&\rho _{M}^{(N)}(\mathbf{v}%
_{1})\left. \mathbf{a}_{1}\cdot \mathbf{v}_{1}+a_{2}\mathbf{v}%
_{1}^{2}+\right.  \notag \\
&&\left. \mathbf{a}_{3}\cdot \mathbf{v}_{1}\mathbf{v}_{1}^{2}+a_{4}\mathbf{v}%
_{1}^{4}+..,\right]  \label{POLY}
\end{eqnarray}%
where the coefficients $a_{i}\equiv $ $a_{i}(\mathbf{r}_{1},t)$ for $i\in
\mathbb{N}
$ are all assumed of order $O(\varepsilon ).$Then validity of the same
equations (\ref{C-1a}),(\ref{C-2a}) and (\ref{C-3a}) requires identically in
$\Omega \times I:$%
\begin{eqnarray}
\int\limits_{U_{1}}d^{3}\mathbf{v}_{1}\delta \rho _{1}^{(N)}(\mathbf{x}%
_{1},t) &=&0 \\
\int\limits_{U_{1}}d^{3}\mathbf{v}_{1}\mathbf{v}_{1}\delta \rho _{1}^{(N)}(%
\mathbf{x}_{1},t) &=&0, \\
\int\limits_{U_{1}}d^{3}\mathbf{v}_{1}\frac{m}{3}\left( \mathbf{v}_{1}-%
\mathbf{V}_{1}(\mathbf{r}_{1},t)\right) ^{2}\delta \rho _{1}^{(N)}(\mathbf{x}%
_{1},t) &=&0.
\end{eqnarray}%
These equations yield linear constraint equations for the coefficients of
the polynomial expansion (\ref{POLY}) of the form:%
\begin{eqnarray}
\int\limits_{U_{1}}d^{3}\mathbf{v}_{1}\rho _{M}^{(N)}(\mathbf{v}_{1})\left[
a_{2}\mathbf{v}_{1}^{2}+a_{4}\mathbf{v}_{1}^{4}+..\right] &=&0  \notag \\
\int\limits_{U_{1}}d^{3}\mathbf{v}_{1}\mathbf{v}_{1}\rho _{M}^{(N)}(\mathbf{v%
}_{1})\left[ \mathbf{a}_{1}\cdot \mathbf{v}_{1}+\mathbf{a}_{3}\cdot \mathbf{v%
}_{1}\mathbf{v}_{1}^{2}+..\right] &=&0,  \notag \\
\int\limits_{U_{1}}d^{3}\mathbf{v}_{1}\frac{m}{3}\mathbf{v}_{1}^{2}\delta
\rho _{1}^{(N)}(\mathbf{x}_{1},t)\left[ a_{2}\mathbf{v}_{1}^{2}+a_{4}\mathbf{%
v}_{1}^{4}+..\right] &=&0.  \label{CONSTRAINT EQS.}
\end{eqnarray}%
It must be stressed that these constraint equations can always be satisfied
in the case MMKE (\ref{MODIFIED MASTER EQ}) is replaced with the homogeneous
Vlasov kinetic equation, namely%
\begin{equation}
L_{1}(\mathbf{F})\rho _{1}^{(N)}(\mathbf{x}_{1},t)=0.
\end{equation}%
However, validity of MMKE and in particular the fact that the Master
collision operator $\mathcal{C}_{1}\left( \rho _{1}^{(N)}|\rho
_{1}^{(N)}\right) $ evaluated in the particular case (\ref{POSITION-1}) is
necessarily non-vanishing for a perturbation $\delta \rho _{1}^{(N)}(\mathbf{%
x}_{1},t)$ of the form (\ref{POLY}) implies that nontrivial relationships
must hold among the coefficients $a_{i}(\mathbf{r}_{1},t)$ for $i\in
\mathbb{N}
.$ Indeed, one can actually show that the same coefficients are necessarily
linearly coupled, requiring
\begin{eqnarray}
a_{2} &=&a_{2}(\mathbf{a}_{1}(\mathbf{r}_{1},t);\mathbf{r}_{1},t);  \notag \\
\mathbf{a}_{3} &=&\mathbf{a}_{3}(a_{2}(\mathbf{r}_{1},t);\mathbf{r}_{1},t);
\notag \\
&&......
\end{eqnarray}%
This implies that the Eqs.(\ref{CONSTRAINT EQS.}) cannot generally be
simultaneously satisfied. The consequence is that validity of Eqs. (\ref%
{C-1a})-(\ref{C-3a}) requires that in Eq.(\ref{POSITION-1}) it should be $%
\delta \rho _{1}^{(N)}(\mathbf{x}_{1},t)\equiv 0.$ Hence under such
constraint conditions it follows that the $1-$body PDF must necessarily
coincide with the kinetic equilibrium PDF of the type indicated above (see
Eq. (\ref{MAXWELLIAN-1})).

\end{document}